\documentclass[
reprint,
  nofootinbib,amsmath,amssymb,aps,preprintnumbers,superscriptaddress,notitlepage
]{revtex4-1}

\usepackage{graphicx}
\usepackage{bm}
\usepackage{braket}
\usepackage{cases} 
\usepackage{here}
\usepackage{color}
\usepackage{upgreek}
\usepackage{comment}
\usepackage{empheq}

\usepackage{hyperref}

\begin{document}

\preprint{KEK-QUP-2023-0011, KEK-TH-2529, KEK-Cosmo-0314, TU-1192}

\title{Probing high frequency gravitational waves with pulsars}

\author{Asuka Ito}
\email[]{asuka.ito@kek.jp}
\affiliation{International Center for Quantum-field Measurement Systems for Studies of the Universe and Particles (QUP), KEK, Tsukuba 305-0801, Japan}
\affiliation{Theory Center, Institute of Particle and Nuclear Studies, KEK, Tsukuba 305-0801, Japan}

\author{Kazunori Kohri}
\affiliation{Theory Center, Institute of Particle and Nuclear Studies, KEK, Tsukuba 305-0801, Japan}
\affiliation{Department of Particles and Nuclear Physics, the Graduate University for Advanced Studies (Sokendai), Tsukuba 305-0801, Japan}
\affiliation{International Center for Quantum-field Measurement Systems for Studies of the Universe and Particles (QUP), KEK, Tsukuba 305-0801, Japan}
\affiliation{Kavli IPMU (WPI), UTIAS, The University of Tokyo, Kashiwa, Chiba 277-8583, Japan}

\author{Kazunori Nakayama}
\email[]{kazunori.nakayama.d3@tohoku.ac.jp}
\affiliation{Department of Physics, Tohoku University, Sendai, Miyagi 980-8578, Japan}
\affiliation{International Center for Quantum-field Measurement Systems for Studies of the Universe and Particles (QUP), KEK, Tsukuba 305-0801, Japan}
\affiliation{Kavli IPMU (WPI), UTIAS, The University of Tokyo, Kashiwa, Chiba 277-8583, Japan}


\begin{abstract}
We study graviton-photon conversion in magnetosphere of a pulsar and explore the possibility of detecting
high frequency gravitational waves with pulsar observations.
It is shown that conversion of one polarization mode of photons can be enhanced significantly due to 
strong magnetic fields around a pulsar.
We also constrain stochastic gravitational waves in frequency range of $10^{8}-10^{9}$\,Hz and 
$10^{13}-10^{27}$\,Hz by using data of observations of the Crab pulsar and the Geminga pulsar. 
Our method widely fills the gap among existing high frequency gravitational wave experiments and boosts
the frequency frontier in gravitational wave observations.
\end{abstract}

\maketitle

%
%
%
%
%
%
%
\section{Introduction}

Detection of gravitational waves from a binary black hole merger with LIGO~\cite{LIGOScientific:2016aoc} 
opened the era of gravitational astronomy/cosmology.
It is important to push forward multi-frequency gravitational wave observations
to investigate our universe~\cite{Kuroda:2015owv}.
In the lower frequency range, we have promising methods for observing gravitational waves like the cosmic microwave observations ($10^{-18}-10^{-16}$\,Hz) and pulsar timing arrays ($10^{-9}-10^{-7}$\,Hz)~\cite{Maggiore:2007ulw,Maggiore:2018sht}.
Indeed, recently, pulsar timing arrays of NANOGrav, PPTA, EPTA, and IPTA detected correlated signals among pulsars~\cite{NANOGrav:2020bcs,Goncharov:2021oub,Chen:2021rqp,Chen:2021ncc}, which may be signals of stochastic gravitational waves.
On the other hand, detection of high frequency gravitational waves above kHz is still under development and 
even new ideas are required~\cite{Aggarwal:2020olq}.
From theoretical viewpoint, there are many sources of high frequency gravitational waves:
scattering of particles in thermal bath~\cite{Ghiglieri:2015nfa,Ghiglieri:2020mhm,Ringwald:2020ist}, inflaton annihilation into gravitons~\cite{Ema:2015dka,Ema:2016hlw,Ema:2020ggo}, bremsstrahlung during the reheating~\cite{Nakayama:2018ptw,Huang:2019lgd,Barman:2023ymn}, preheating after inflation~\cite{Khlebnikov:1997di,Easther:2006gt,Garcia-Bellido:2007nns},
decay of heavy particles into the graviton pair~\cite{Ema:2021fdz}.
Thus the detection of high frequency gravitational waves provides us with rich information on the fundamental physcis models.

One natural direction to seek methods of detecting high frequency gravitational waves is to utilize tabletop experiments, 
since gravitational wave detectors often become sensitive when its size is comparable to wavelength of gravitational waves.
For example, the magnon gravitational wave detector utilizing resonant excitation of magnons by gravitational waves
was proposed for detecting gravitational waves around GHz~\cite{Ito:2019wcb,Ito:2020wxi,Ito:2022rxn}.
Gravitational waves can also be converted into photons under the background magnetic field~\cite{gertsenshtein1962wave,Raffelt:1987im}. Along this direction, new high frequency gravitational wave detection methods with the use of axion detection experiments have been proposed intensively~\cite{Ejlli:2019bqj,Ringwald:2020ist,Aggarwal:2020olq,Berlin:2021txa,Domcke:2022rgu,Tobar:2022pie}.
Another possible way is to utilize astrophysical observations of photons with various frequencies.
In Refs.~\cite{Dolgov:2012be,Domcke:2020yzq} it has been proposed that the observation of microwave/X-ray photons gives constraint on the stochastic gravitational waves at the corresponding frequency, depending on the strength of the primordial magnetic field.\footnote{
    The inverse process, i.e., the cosmic background photon conversion into gravitons has also been investigated~\cite{Chen:1994ch,Cillis:1996qy,Pshirkov:2009sf,Chen:2013gva,Fujita:2020rdx}.
}
In this letter, we propose a new detection method of high frequency gravitational waves with observations of pulsars.

Pulsars are extremely fast rotating neutron stars that originated from
supernova explosions and possess strong magnetic fields with  $\mathcal{O}$($10^{12}$) G~\cite{Meszaros:1992tf}.~\footnote{Here, for simplicity we do not consider magnetars which have stronger magnetic field than the critical value $\sim 4 \times 10^{13}$G because of the nonlinear QED effects (e.g., see Refs.~\cite{Adler:1971wn,Kohri:2001wx}) .}
This strong magnetic field accelerates electrons to energies of approximately
$\mathcal{O}$(1) TeV, and they have a power-law energy spectrum starting at
$\mathcal{O}$(1) MeV and a cutoff structure at the energy of $\mathcal{O}$(1) TeV.  It is known that photons with pulsed and stationary
components are produced by inverse Compton scattering, synchrotron
radiation, and curvature radiation by these high-energy electrons.
Here, we propose that photons converted from the background gravitational wave by the magnetic field may be mixed in the observed photon signals. Then, if we require the component converted from the background gravitational wave not to exceed the total signals of the observed photons, we show that the observational data of photons emitted from pulsars conservatively provide upper bounds
on the characteristic strain for the background gravitational wave to be
$h_c < 10^{-26} - 10^{-14}$ at frequencies from $10^8$ Hz to $10^{27}$
Hz.

\section{Photon propagation in magnetized plasma}
\label{disper}
Let us derive the modified dispersion relation of photons in magnetized plasma.
Around a pulsar, there exists magnetic fields and charged particles such as electrons and protons.
We consider cold plasma, namely it's thermal motion is negligible.
When electromagnetic fields propagate in plasma medium, charged particles are fluctuated.
A charged particle with a mass $m_{(i)}$ and a charge $e_{(i)}$ ($i$ specifies species) obeys the equation:
\begin{equation}
  m_{(i)} \frac{\partial \bm{v}_{(i)}}{\partial t} = e_{(i)}\left( \bm{E} +  \bm{v}_{(i)} \times \bar{\bm{B}} \right), \label{v}
\end{equation}
where $\bar{\bm{B}}$ represetns the background magnetic field. 
The velocity of the charaged particle $\bm{v}_{(i)}$ and the electric field $\bm{E}$ are treated as perturbations.
Accordingly, we have electric current
\begin{equation}
  \bm{j} = e n_{(i)} \bm{v}_{(i)} .
\end{equation}
where $n_{(i)}$ is the number density of the charged particle.
Then the maxwell equations for perturbations are given by
\begin{empheq}[left=\empheqlbrace]{align} 
  {\rm rot} \bm{E} &= - \frac{\partial \bm{B}}{\partial t} ,  \\
  {\rm rot} \bm{B} &=  \bm{j} + \frac{\partial \bm{E}}{\partial t} . \label{maxwell}
\end{empheq}
We will only consider a background magnetic field perpendicular to the propagation direction of photons, 
because only such a configuration contributes to graviton-photon conversion.
Then, without loss of generality, 
we take the direction of the magnetic field and of the propagation of photons along $y$-axis and $z$-axis, respectively.
Assuming a function form of $(\bm{E}, \bm{B}, \bm{v}_{(i)}) \propto e^{-i (\omega t - \bm{k}\cdot \bm{x})}$,
from Eqs.\,(\ref{v})-(\ref{maxwell}), one can deduce a following dispersion relation,
\begin{widetext}
\begin{equation}
  \begin{pmatrix}
                 1 - \frac{\omega_{p,(i)}^{2}}{\omega^{2}-\omega_{c,(i)}^{2}} 
               - \frac{k^{2}}{\omega^{2}}  
               &  0
               &  +i \frac{\omega_{c,(i)}}{\omega}\frac{\omega_{p,(i)}^{2}}{\omega^{2}-\omega_{c,(i)}^{2}} \\
                0
              & 1 - \frac{\omega_{p,(i)}^{2}}{\omega^{2}} 
                   -  \frac{k^{2}}{\omega^{2}}             
               & 0  \\
                 -i \frac{\omega_{c,(i)}}{\omega}\frac{\omega_{p,(i)}^{2}}{\omega^{2}-\omega_{c,(i)}^{2}}
              &  0  
              &  1 - \frac{\omega_{p,(i)}^{2}}{\omega^{2}-\omega_{c,(i)}^{2}}
              
  \end{pmatrix}
\begin{pmatrix}
                     A_{x} \\
                     A_{y} \\
                     A_{z}
  \end{pmatrix}
            = 0 .   \label{modis}
\end{equation}
\end{widetext}
%
%
%
%
In the expression, rewrote the electric field 
by the vector potential $\bm{A}$, i.e., $\bm{E} = - \partial_{t} \bm{A} = i\omega \bm{A}$.%
\footnote{We neglect the scalar potential, if any, because we are interested in only propagation degree of freedom of 
electromagnetic fields.
}
The plasma frequency and the cyclotron frequency are defined by 
$\omega_{p,(i)} = \sqrt{\frac{4\pi\alpha n_{(i)}}{m_{(i)}}}$ and 
$\omega_{c,(i)} = \frac{e\bar{B}}{m_{(i)}}$, respectively.
\begin{figure}[ht]
\centering
\includegraphics[width=6cm]{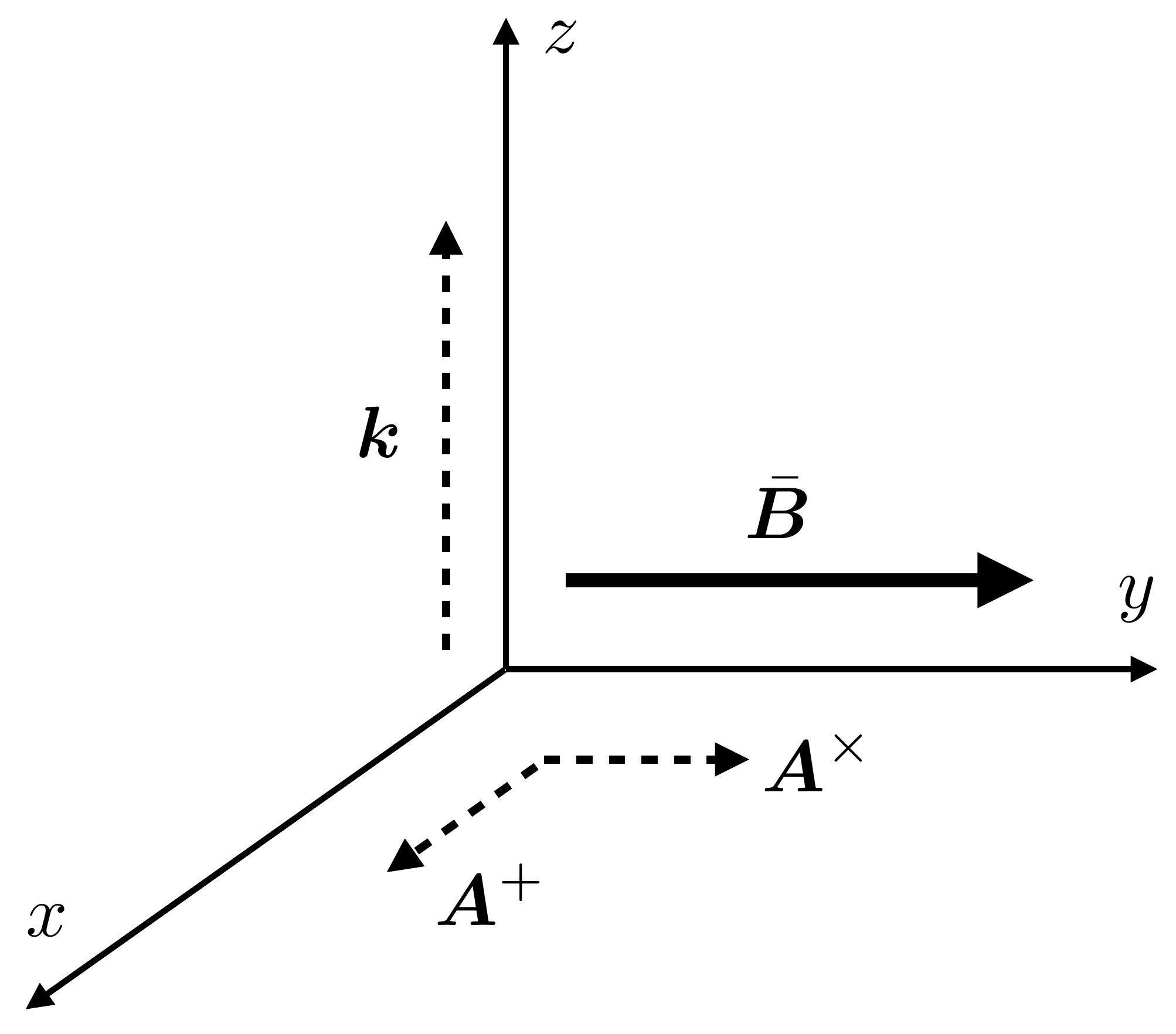}
\caption{The configuration of linear polarization modes to the propagation direction $\bm{k}$ and the 
the background magnetic field $\bar{\bm{B}}$ is shown.} \label{Csekibun}
\end{figure}
\section{Graviton-photon conversion}
We now consider graviton-photon conversion around a pulsar.
We first consider the mixing between gravitons and photons in vacuum and promote the result into
magnetized plasma background later.
The action of photons in QED is
\begin{eqnarray}
    S &=& \int d^{4}x \sqrt{-g}\bigg[ \frac{M_{{\rm pl}}^{2}}{2}R
                           -\frac{1}{4} F_{\mu\nu}F^{\mu\nu}  \nonumber \\
      &\ &   + \frac{\alpha^{2}}{90 m_{e}^{4}} 
          \left( (F_{\mu\nu}F^{\mu\nu})^{2} + \frac{7}{4}(F_{\mu\nu} \tilde{F}^{\mu\nu})^{2} \right)
                          \bigg] ,  \label{ac}
\end{eqnarray}
where $M_{{\rm pl}}$ represents the reduced Planck mass, $R$ is the Ricci scalar,
$g$ is the determinant of a metric $g_{\mu\nu}$, 
$\alpha$ is the fine structure constant, and $m_{e}$ is the electron mass.
The field strength of electromagnetic fields is defined by 
$F_{\mu\nu} = \partial_{\mu}\mathcal{A}_{\nu} - \partial_{\nu}\mathcal{A}_{\mu}$ where $\mathcal{A}_{\mu}$ is the vector potential.
$\tilde{F}^{\mu\nu} = - 1/2 \epsilon^{\mu\nu\rho\sigma}F_{\rho\sigma}$ ($\epsilon^{0123} = -1$) is the dual of the field strength.
The third term is the Euler-Heisenberg term from the vacuum polarization~\cite{Heisenberg:1936nmg}. 
We now expand the vector potential and the metric as
\begin{empheq}[left=\empheqlbrace]{align} 
  \mathcal{A}_{\mu}(x) &= \bar{A}_{\mu} + A_{\mu}(x) , \\
  \label{met033}
  g_{\mu\nu}(x) &= \eta_{\mu\nu} + \frac{2}{M_{{\rm pl}}} h_{\mu\nu}(x) .       
\end{empheq}
Here, $\bar{A}_{\mu}$ consists of background magnetic fields $\bar{B}^{i} = \epsilon^{ijk} \partial_{j} \bar{A}_{k}$
 around a pulsar, 
$\eta_{\mu\nu}$ stands for the Minkowski metric, and 
$h_{\mu\nu}(x)$ is a traceless-transverse tensor representing gravitational waves.
Below we take the gauge $A_0=0$. We will only consider the transverse mode, $A_x$ and $A_y$ as shown in Fig.~\ref{Csekibun}, and neglect their mixing with $A_z$ in Eq.~(\ref{modis}) since the mixing is small in the case of our interest.

In the presence of background magnetic fields, one can expand the action (\ref{ac}) at
the second order of perturbations,
\begin{widetext}
\begin{eqnarray}
    \delta S^{(2)} &=& \int d^{4}x \bigg[ -\frac{1}{2} \left( \partial_{\mu}h_{ij} \right)^{2} 
                                    - \frac{1}{2} \left( \partial_{\mu}A_{i}\right)^{2}
                                    + \frac{2}{M_{{\rm pl}}} \epsilon_{ijk} \bar{B}^{k} h^{jl} \partial_{i}A^{l} \nonumber \\
   &\ & + \frac{\alpha^{2}}{90 m_{e}^{4}} \bigg(   
         16 \bar{B}^{i}\bar{B}^{j} \Big( \delta_{ij}(\partial_{k}A_{l})^{2}
                                         - (\partial_{k}A_{i})(\partial_{k}A_{j}) 
                                         - (\partial_{i}A_{k})(\partial_{j}A_{k}) \Big) 
                                       + 28 \Big((\partial_{0}A_{i})\bar{B}_{i}\Big)^{2}   \bigg)
                          \bigg] . \nonumber \\  \label{ac2}
\end{eqnarray}
\end{widetext}
where 
higher order terms of $\alpha$ have been dropped.
The third term represents the mixing between gravitons and photons due to the background magnetic field.
Note that only magnetic fields perpendicular to the propagation direction of gravitons and/or photons contribute to the mixing.
Our configurations are shown in Fig.~\ref{Csekibun}. We consider gravitons and/or photons propagating along $z$-direction and the background magnetic field orthogonal to the propagation direction, which is taken to be $y$-direction, $\bar{\bm{B}}=(0, \bar{B}, 0)$.
One can also choose the polarization bases for the vector and the tensor as
\begin{eqnarray}
  &e^{+}_{i} =  \begin{pmatrix}
                     1 \\
                     0 \\
                     0
                    \end{pmatrix} , \quad
  e^{\times}_{i} =  \begin{pmatrix}
                     0 \\
                     1 \\
                     0
                    \end{pmatrix} , \quad  \nonumber \\
  &\epsilon^{+}_{ij}= \frac{1}{\sqrt{2}} \begin{pmatrix}
              1 & 0 & 0\\
              0 & -1 & 0\\
              0 & 0 & 0
              \end{pmatrix} ,  \quad
  \epsilon^{\times}_{ij}= \frac{1}{\sqrt{2}} \begin{pmatrix}
              0 & 1 & 0\\
              1 & 0 & 0\\
              0 & 0 & 0
              \end{pmatrix} .   \label{basis}
\end{eqnarray}
Using the bases (\ref{basis}), the electromagnetic field and the gravitational wave can be expanded as follows:
\begin{empheq}[left=\empheqlbrace]{align} 
  A_{i} = e^{-i(\omega t-kz)} A^{\sigma}(z) e^{\sigma}_{i} , \\
  h_{ij} = e^{-i(\omega t-kz)} h^{\sigma}(z) \epsilon_{ij}^{\sigma} ,     \label{planar}
\end{empheq}

We can now derive coupled equations of motion for the photon 
and the gravition of each polarization modes.
Since there exists ionized particles around a pulsar,
we also take into account the modification to the dispersion relation of photons (\ref{modis}).
From Eqs.\,(\ref{ac2})-(\ref{planar}) and (\ref{modis}), we obtain
%
%
%
\begin{equation}
  \left[ i \partial_{z} 
         +
                 \begin{pmatrix}
              -\frac{1}{2\omega} \frac{\omega^{2}\omega_{p,(i)}^{2}}{\omega^{2}-\omega_{c,(i)}^{2}}  
              + \frac{1}{2\omega} \frac{16\alpha^{2}\bar{B}^{2}\omega^{2}}{45m_{e}^{4}}
              & i \frac{B}{\sqrt{2}M_{{\rm pl}}} \\
              -i \frac{B}{\sqrt{2}M_{{\rm pl}}} 
              & 0
              \end{pmatrix}
      \right] 
      \begin{pmatrix}
                     A^{+}(z) \\
                     h^{+}(z) 
      \end{pmatrix}
            \simeq 0 ,   \label{mat}
\end{equation}
\begin{equation}
  \left[ i \partial_{z} 
         +
                 \begin{pmatrix}
             - \frac{\omega_{p,(i)}^{2}}{2\omega}  
             + \frac{1}{2\omega} \frac{28\alpha^{2}\bar{B}^{2}\omega^{2}}{45m_{e}^{4}}  
              & i \frac{B}{\sqrt{2}M_{{\rm pl}}} \\
              -i \frac{B}{\sqrt{2}M_{{\rm pl}}} 
              & 0
              \end{pmatrix}
      \right] 
      \begin{pmatrix}
                     A^{\times}(z) \\
                     h^{\times}(z) 
      \end{pmatrix}
            \simeq 0 .   \label{mat2}
\end{equation}
As deriving the equations, 
we have 
assumed that the scale of conversion between photons and gravitons is much longer than $k^{-1}$ and photons
are ultrarelativistic, i.e., $\omega \simeq k$.
Also, we neglected spatial dependence of $\bar{B}$ by considering enough small region $\Delta r$
where $\bar{B}$ can be regarded 
as a constant. 
since it is not of our interest.
By diagonalizing the matrix in Eqs.\,(\ref{mat}) and (\ref{mat2}), one can solve the equations and obtain the conversion rate
between electromagnetic fields and gravitational waves of the plus modes:
\begin{widetext}
\begin{equation}
  P^{(+)}_{(i)}(A^{+} \leftrightarrow h^{+}) = 
      \frac{\frac{8 \bar{B}^{2}\omega^{2}}{M_{{\rm pl}}^{2}}}
      {\left( \frac{\omega^{2}\omega_{p,(i)}^{2}}{\omega^{2}-\omega_{c,(i)}^{2}} 
            - \frac{16\alpha^{2}\bar{B}^{2}\omega^{2}}{45m_{e}^{4}} \right)^{2}
      + \frac{8 \bar{B}^{2}\omega^{2}}{M_{{\rm pl}}^{2}} }
      \times
      \sin^{2}\left( \frac{\sqrt{\left( \frac{\omega^{2}\omega_{p,(i)}^{2}}{\omega^{2}-\omega_{c,(i)}^{2}} 
                                        - \frac{16\alpha^{2}\bar{B}^{2}\omega^{2}}{45m_{e}^{4}} \right)^{2} 
                                        + \frac{8 \bar{B}^{2}\omega^{2}}{M_{{\rm pl}}^{2}} }}{4\omega} \Delta r
                                        \right) , \label{cp}
\end{equation}
\end{widetext}
and of the cross modes:
\begin{widetext}
\begin{equation}
  P^{(\times)}_{(i)}(A^{\times} \leftrightarrow h^{\times}) = 
      \frac{\frac{8 \bar{B}^{2}\omega^{2}}{M_{{\rm pl}}^{2}}}
      {\left( \omega_{p,(i)}^{2} - \frac{28\alpha^{2}\bar{B}^{2}\omega^{2}}{45m_{e}^{4}} \right)^{2}
      + \frac{8 \bar{B}^{2}\omega^{2}}{M_{{\rm pl}}^{2}} }
      \times
      \sin^{2}\left( \frac{\sqrt{\left( \omega_{p,(i)}^{2} 
                                        - \frac{28\alpha^{2}\bar{B}^{2}\omega^{2}}{45m_{e}^{4}} \right)^{2} 
                                        + \frac{8 \bar{B}^{2}\omega^{2}}{M_{{\rm pl}}^{2}} }}{4\omega} \Delta r
                                        \right) .  \label{cc}
\end{equation}
\end{widetext}
It is worth noting that there is the contribution of $\omega_{c,(i)}$ in the conversion rate of the plus mode, while
there is not for the cross mode.
In the next section, we will see that the contribution gives rise to big difference of the conversion rate between 
the polarization.
We will also evaluate photon flux from conversion of gravitational waves around a pulsar and give 
constraints on high frequency gravitational waves by using observed photon spectrum.
\section{Photon flux from gravitational waves around a pulsar}
We now estimate photon flux converted from gravitational waves around a pulsar.
To this end, based on the Goldreich-Julian model~\cite{Goldreich:1969sb}, we simply model the 
magnetosphere of a pulsar; magnetic fields has a dipole like distribution: 
\begin{equation}
  \bar{B}(r) = \sqrt{\frac{2}{3}} \bar{B}_{0} \left( \frac{r}{r_{0}} \right)^{-3},
\end{equation}
where $\bar{B}_{0}$ is an amplitude of the magnetic field at the surface of a neutron star with a radius $r_{0}$.
We took an average of the direction of magnetic fields by assuming equipartition, so that
the component of magnetic fields perpendicular to poropagation direction of photons and gravitons obtained 
the factor of $\sqrt{2/3}$.
Inspired by the Goldreich-Julian model~\cite{Goldreich:1969sb}, the number density of electrons or protons is parametrized 
as follows,
\begin{equation}
  n_{(i)}(r) = 7\times 10^{-2} \times \left( \frac{1 {\rm s}}{T}  \right) 
               \left( \frac{\bar{B}(r)}{1 {\rm G}}  \right) {\rm cm}^{-3} .
    \label{ne}
\end{equation}
Here $T$ is the rotation period of a neutron star.

We first show that the amplitude of the conversion rate of the plus mode (\ref{cp}) is always higher than that of the cross mode (\ref{cc}).
Around a pulsar, one can estimate the each relevant parameter as follows:
\begin{widetext}
\begin{eqnarray}
  \omega_{p,(i)} &=& 1.5 \times 10^{11} \times \left( \frac{511 \, {\rm keV}}{m_{i}} \right)^{1/2} 
                                      \left( \frac{\bar{B}}{ 10^{12}\,  {\rm G}} \right)^{1/2}
                                      \left( \frac{10 \, {\rm ms}}{T} \right)^{1/2} {\rm Hz} ,  \label{pla}  \\
  \omega_{c,(i)} &=& 1.8 \times 10^{19} \times \left( \frac{511 \, {\rm keV}}{m_{i}} \right) 
                                      \left( \frac{\bar{B}}{10^{12}\,  {\rm G}} \right)
                                       {\rm Hz} , \label{cy} \\
  \omega_{{\rm QED}} &\equiv& \frac{\alpha\bar{B}\omega}{m_{e}^{2}} 
                  = 3.4 \times 10^{16} \times 
                                      \left( \frac{\bar{B}}{ 10^{12}\,  {\rm G}} \right)
                                      \left( \frac{\omega/2\pi}{ 10^{19}\,  {\rm Hz}} \right)
                                       {\rm Hz} ,  \label{qed}  \\           
  \Omega &\equiv& \sqrt{\frac{8 \bar{B}\omega}{M_{{\rm pl}}}}
                  = 2.5 \times 10^{9} \times 
                                      \left( \frac{\bar{B}}{10^{12}\,  {\rm G}} \right)^{1/2}
                                      \left( \frac{\omega/2\pi}{ 10^{19}\,  {\rm Hz}} \right)^{1/2}
                                       {\rm Hz} ,       \label{coup}                                      
\end{eqnarray}
\end{widetext}
where $\omega_{{\rm QED}}$ represents the magnitude of the nonlinear QED effect and $\Omega$ is the coupling strength
of the graviton-photon mixing.
First of all, from Eqs.\,(\ref{pla})-(\ref{qed}), one can find that the term 
$\sqrt{\frac{\omega^{2}\omega_{p,(i)}^{2}}{\omega^{2}-\omega_{c,(i)}^{2}}}$, which appeared
in Eq.\,(\ref{cp}), is always much smaller than $\omega_{{\rm QED}}$ for 
typical values of the parameters characterizing magnetosphere around a pulsar.
Note that this is true not only for electrons but also for protons.
Then, we can neglect the irrelevant term and approximate the conversion rate of the plus mode as
\begin{widetext}
\begin{equation}
  \  P^{(+)}(r,\omega) \simeq 
      \frac{\frac{8 \bar{B}^{2}\omega^{2}}{M_{{\rm pl}}^{2}}}
      {\left( 
             \frac{16\alpha^{2}\bar{B}^{2}\omega^{2}}{45m_{e}^{4}} \right)^{2}
      + \frac{8 \bar{B}^{2}\omega^{2}}{M_{{\rm pl}}^{2}} }
      \times
      \sin^{2}\left( \frac{\sqrt{\left( 
                                         \frac{16\alpha^{2}\bar{B}^{2}\omega^{2}}{45m_{e}^{4}} \right)^{2} 
                                        + \frac{8 \bar{B}^{2}\omega^{2}}{M_{{\rm pl}}^{2}} }}{4\omega} \Delta r
                                        \right) .  \label{cp20}
\end{equation}
\end{widetext}
Notably, when $\omega_{p,(i)} > \omega_{{\rm QED}}$, the amplitude of the above conversion rate is significantly higher than that of 
the cross mode (\ref{cc}).
Even when $\omega_{p,(i)} < \omega_{{\rm QED}}$, the amplitude of Eq.\,(\ref{cp20}) is higher than that of Eq.\,(\ref{cc})
by a numerical factor.
Thus hereafter, we will only consider the conversion of the plus mode.%
\footnote{In small parameter region which satisfies $\omega_{p,(i)} \simeq \sqrt{28/45}\omega_{{\rm QED}}$, 
the amplitude of the conversion rate of the cross mode can be of order unity.
However, since whether such resonance occurs or not highly depends on the model of magnetosphere, 
we do not consider such a possibility to give a conservative result.
}
It should be noticed that, when we consider graviton-photon conversion around a pulsar, 
we need to take into account $r$ dependence of the background magnetic field. Thus we cannot simply use the formula (\ref{cp20}).
To illustrate this, let us suppose that, starting from the pure graviton state at $r\to -\infty$, the magnetic field is adiabatically turned on and peaked around $r\sim 0$ and then adiabatically turned off at $r\to\infty$. Then the probability to find the photon at $r\to\infty$ vanishes.  
In reality, however, the oscillation length is longer than $R$
($R$ is the end of the magnetosphere, which is usually characterized by the light cylinder $R=T/2\pi$)
for $f\lesssim 10^{22}\,{\rm GHz}$ hence the adiabaticity is violated within $|r|<R$. Thus the graviton-photon conversion takes place in the pulsar magnetosphere. For higher frequency, it is not very clear which fraction of gravitons are converted to photons after passing through the magnetosphere. One possibility is that the conversion happens at the boundary of the light cylinder, at which the change of the magnetic field is rather sudden. Still, for very high frequency, the oscillation length becomes so small that any spatial variation of the magnetic field is regarded as adiabatic.
In the following we numerically integrate Eq.\,(\ref{mat}) to calculate the conversion rate 
$P^{(+)}_{r_0 - R}$ when gravitons propagate from $r_0$ to $R$.
For $f\lesssim 10^{22}\,{\rm GHz}$, this gives a reasonable estimate of the probability to find photons at the detector far from the pulsar. For higher frequency, one should be cautious about the derived constraint.

Now let us estimate flux of the photons converted from stochastic gravitational waves and 
show the ability of pulsars as gravitational wave detectors.
Around a pulsar, there would exist stochastic gravitational waves, which can be characterized with the characteristic amplitude
defined by~\cite{Maggiore:1999vm}
\begin{equation}
    \langle h_{ij}(t)h^{ij}(t)\rangle=2\int_{f=0}^{f=\infty}d(\log f) h^{2}_{c}(f) . \label{hcteigi}
\end{equation}
Due to graviton-photon conversion in magnetosphere of a pulsar, photons are generated.
Its flux $F$ is given by
\begin{equation}
  F = \frac{2\pi^2 M_{\rm pl}^{2} f h_{c}^{2}R^2}{ d^{2} } P^{(+)}_{r_0 - R}
\end{equation}
where $d$ is the distance between a pulsar and the Earth.

\begin{figure}[H]
\centering
\includegraphics[width=8.8cm]{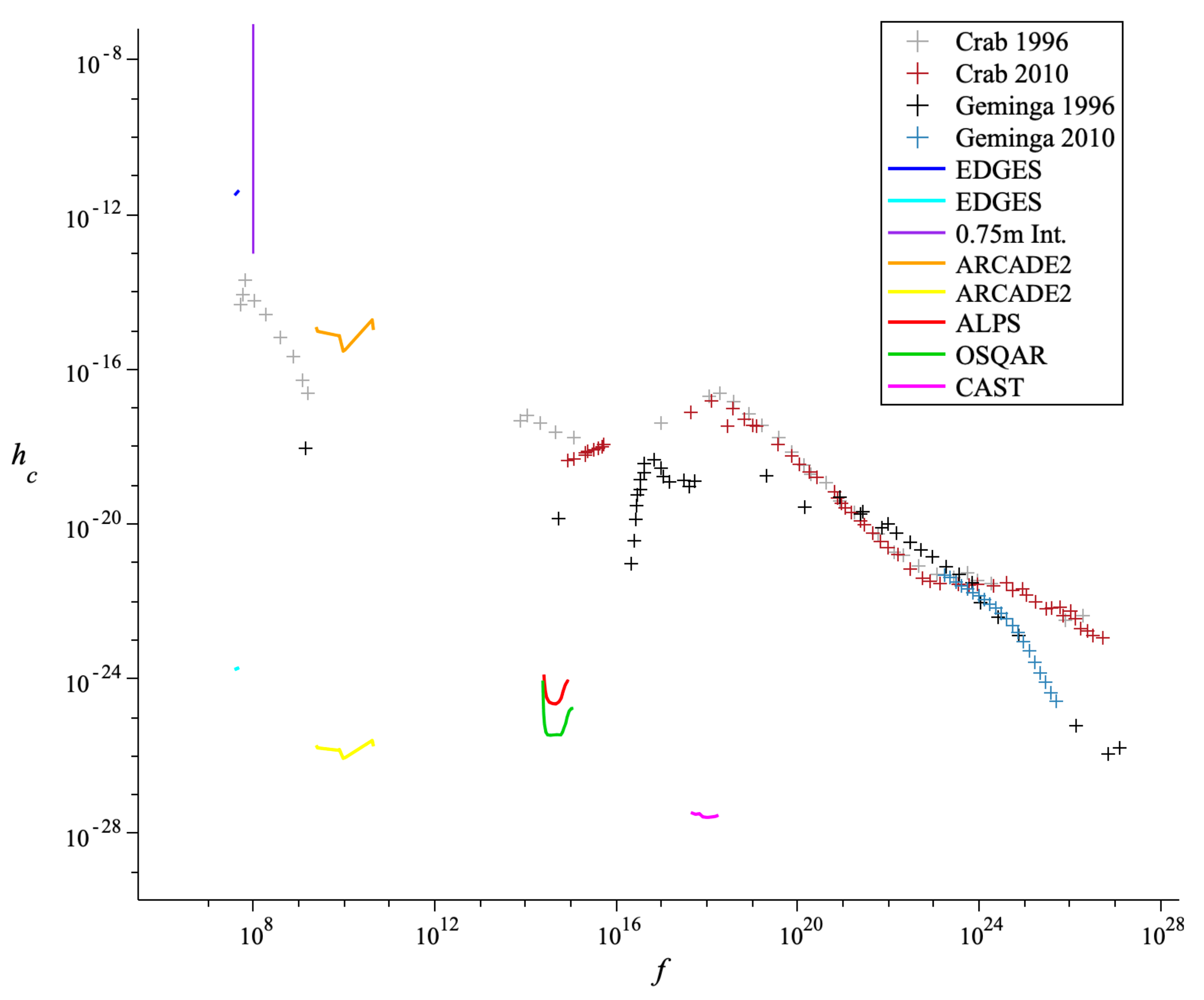}
\caption{Our constraints on $h_{c}$ of stochastic gravitational waves for frequency $f\,{\rm [Hz]}$ from the Crab pulsar observations in 
1996s~\cite{1996ASPC..105..307T}, 2010s~\cite{2010AA...523A...2M}, and from the Geminga pulsar 
observations in 1996~\cite{1996ASPC..105..307T}, 2010~\cite{Fermi-LAT:2010mou} 
are depicted with the gray, brawn, black, and navy points, respectively.
The blue (cyan) and the orange (yellow) lines respectively represents 
the constraints with EDGES and ARCADE2 for maximal (minimum) amplitude of cosmological magnetic fields~\cite{Domcke:2020yzq}.
The violet line is the upper limit from 0.75~m interferometer~\cite{Akutsu:2008qv}.
The red, green, and pink lines represents constraints with ALPS, OSQAR, and CAST, respectively~\cite{Ejlli:2019bqj}.
} \label{hc}
\end{figure}

We use observed spectra of the Crab pulsar and 
the Geminga pulsar~\cite{1996ASPC..105..307T,2010AA...523A...2M,Fermi-LAT:2010mou,Buhler:2013zrp}
to give upper limits on stochastic gravitational waves.
The parameters charactrizing the pulsars are 
$r_{0} = 10 \,{\rm km}$, $\bar{B}_{0} = 7.6\times 10^{12} \,{\rm G}$, $T = 33\,{\rm ms}$,  
$d = 2.0 {\rm \,kpc}$ for the Crab pulsar and 
$r_{0} = 10 \,{\rm km}$, $\bar{B}_{0} = 1.6\times 10^{12} \,{\rm G}$, $T = 237\,{\rm ms}$,  
$d = 0.25 {\rm \,kpc}$ for the Geminga pulsar.
The result is shown in Fig.\,\ref{hc}.
There is no observation of spectra between $10^{9}-10^{13}$\,Hz due to absorption by atmosphere.
Nevertheless, one sees that our new constraints widely fill the gap among the existing experiments and 
boost the highest observable frequency.
The upper limits from EDGES and ARCADE2 have large uncertainty depending on the amplitude of cosmological magnetic fields.
Our constraints around $10^{8}$\,Hz are in the middle of the uncertainty.
There are constraints from ALPS and OSQAR around $10^{15}$\,Hz, and CAST around $10^{18}$\,Hz~\cite{Ejlli:2019bqj}.
Their limits around the frequency regions are still stronger than ours.
In the higher frequency region above $10^{18}$\,Hz where there has not been any constraints, we 
put new limits up to $10^{27}$\,Hz.

\section{Conclusion}
We studied graviton-photon conversion in magnetosphere of a pulsar.
It turned out that
graviton-photon conversion can be significantly effective for photons of a polarization mode perpendicular to magnetic fields,
which is called plus mode in this letter,
compared to a polarization mode parallel to magnetic fields (cross mode) due to 
the large magnetic field around a pulsar.
This enhancement of graviton-photon conversion rate does not happens for typical magnetic fields in our universe
such as cosmological magnetic fields background~\cite{Raffelt:1987im}.
It is also noted that the enhancement is absent for axion-photon conversion where 
only a polarization mode parallel to magnetic fields (cross mode) of photons are mixed with axions.
Therefore, it is characteristic only for graviton-photon conversion in existence of strong magnetic fields and plasma.

We also demonstrated ability of pulsar observations as high frequency gravitaitonal wave detectors by giving constraints on
stochastic gravitaitonal waves in frequency range from $10^{8}$\,Hz to $10^{9}$\,Hz and from $10^{13}$\,Hz to $10^{27}$\,Hz 
with data of observations of the Crab pulsar and the Geminga pulsar.
As one can see from Fig.\,\ref{hc},
our method enables us to fill the gap among exsisting high frequency gravitaional wave observations.
Moreover, the frequency frontier in gravitational wave observations is significantly extended from 
$10^{18}$\,Hz to $10^{27}$\,Hz.%
\footnote{Recently, \cite{Liu:2023mll} appeared on arXiv, which also considered graviton-photon conversion
around magnetosphere of planets.
They gave constraints around $10^{11}$\,Hz$-10^{15}$\,Hz and
$10^{17}$\,Hz$-10^{23}$\,Hz.
}
\begin{acknowledgments}
This work was supported by World Premier International Research Center Initiative (WPI), MEXT, Japan.
A.\,I.\ was in part supported by JSPS KAKENHI Grant Numbers JP21J00162, JP22K14034 (A.I.), and MEXT KAKENHI Grant Numbers JP22H05270 (K.K.).
\end{acknowledgments}

\bibliography{pulsar}

\end{document}